\documentclass[acmsmall]{acmart}
\usepackage{float}
\usepackage{graphicx}
\usepackage{multirow}
\usepackage{pgfplots}

\AtBeginDocument{%
  }

\copyrightyear{2025}
\acmYear{2025}
\acmDOI{XXXXXXX.XXXXXXX}

\acmConference[FSE2025]{The ACM International Conference on the Foundations of Software Engineering}{June 23--27, 2025}{Trondheim, Norway}
\acmISBN{978-1-4503-XXXX-X/18/06}

\begin{document}

\title{An Empirical Study on Self-correcting Large Language Models for Data Science Code Generation}

\author{Thai Tang Quoc}
\email{tqthai.sdh222@hcmut.edu.vn}
\author{Duc Ha Minh}
\email{hmduc.sdh211@hcmut.edu.vn}
\author{Tho Quan Thanh}
\email{qttho@hcmut.edu.vn}
\affiliation{%
  \institution{Ho Chi Minh City University of Technology}
  \city{Ho Chi Minh City}
  \country{Vietnam}
}

\author{Anh Nguyen-Duc}
\affiliation{%
  \institution{University of South Eastern Norway}
  \city{B{\o} i Telemark}
  \country{Norway}}
\email{anh.nguyen.duc@usn.no}

\renewcommand{\shortauthors}{Thai et al.}

\begin{abstract}
Large Language Models (LLMs) have recently advanced many applications on software engineering tasks, particularly the potential for code generation. Among contemporary challenges, code generated by LLMs often suffers from inaccuracies and hallucinations, requiring external inputs to correct. One recent strategy to fix these issues is to refine the code generated from LLMs using the input from the model itself (self-augmented). In this work, we proposed a novel method, namely CoT-SelfEvolve. CoT-SelfEvolve iteratively and automatically refines code through a self-correcting process, guided by a chain of thought constructed from real-world programming problem feedback.  Focusing on data science code, including Python libraries such as NumPy and Pandas, our evaluations on the DS-1000 dataset demonstrate that CoT-SelfEvolve significantly outperforms existing models in solving complex problems. The framework shows substantial improvements in both initial code generation and subsequent iterations, with the model's accuracy increasing significantly with each additional iteration. This highlights the effectiveness of using chain-of-thought prompting to address complexities revealed by program executor traceback error messages. We also discuss how CoT-SelfEvolve can be integrated into continuous software engineering environments, providing a practical solution for improving LLM-based code generation.
\end{abstract}



\received{XXX}
\received[revised]{XXY}
\received[accepted]{XXZ}

\maketitle

\section{Introduction}
Code generation aims to automatically produce source code based on given specifications or requirements, which enables developers to save time by reducing manual implementation efforts and allows them to focus on more innovative activities \cite{dehaerne_code_2022}. The recent advancements in Large Language Models (LLMs) have significantly advanced this area \cite{chen2021evaluating,thoppilan2022lamda,achiam2023gpt}. LLMs or Foundation Models (FMs)s are Artificial Intelligence models built with extensive pre-trained corpus, leveraging Deep Learning (DL) techniques, often neural network architectures, to process and produce natural language. LLMs like Codex~\cite{chen2021evaluating}, LaMDA~\cite{thoppilan2022lamda}, and GPT-4~\cite{achiam2023gpt} are showcasing remarkable proficiency in generating high-quality code across multiple programming languages, understanding complex code structures, and even translating natural language specifications into functional code. However, code generated by LLMs can often be inaccurate or exhibit hallucinations, necessitating tasks closely related to code debugging and software repair.

LLM-based assisting tools for data scientists are important but face several challenges. Firstly, data science code, which involves exploratory tasks like data analysis, model building, visualization, and deployment, is particularly susceptible to errors and bugs. The complexity of these tasks requires models to understand domain-specific libraries (e.g., NumPy, Pandas, Scikit-learn) and workflows. Secondly, while LLMs can suggest code snippets for integration into a developer's codebase, the effectiveness of these suggestions largely depends on the developers themselves \cite{bird_taking_2023,moradi_dakhel_github_2023}. Experienced developers can effectively discern and refine these suggestions, but novice programmers might struggle, leading to potential misunderstandings or improper implementations \cite{moradi_dakhel_github_2023}.

To address the second challenge, recent research has begun to explore the concept of "self-correcting" LLMs, where the models iteratively improve upon their own outputs through feedback loops and refinement processes~\cite{jiang2023selfevolve, schick2024toolformer, chen2023teaching}. The process begins with an AI model generating an initial code snippet based on a given input or prompt. Once the initial code is generated, it undergoes evaluation, which identifies errors, or areas for improvement in the generated code. The AI model incorporates the feedback from the evaluation step into its learning process. This not only refines the specific code snippet but also improves its overall understanding and capability for future code-generation tasks. 

The feedback from the evaluation step is used to refine the code. The AI model iterates on its initial output, making adjustments and improvements. This process may involve re-generating parts of the code, optimizing algorithms, fixing bugs, or enhancing readability. Some methods leverage unit tests, code execution traces, or formal verification techniques to provide the evaluation to the model, guiding it toward a correct solution. Others utilize reinforcement learning techniques to train the LLM to generate code more likely to pass predefined test cases. While these approaches show some promise, they often struggle with complex debugging scenarios, especially those involving subtle logical errors or intricate dependencies within the code.

We utilize an external knowledge base by extracting conversations from developers' forums, i.e., StackOverflow, to construct a domain-relevant knowledge base for guiding LLMs. The assumption is that structured conversation regarding bug fixing  By adopting Chain of Thought (CoT) patterns, the learning process of LLMs can mimic the way developers discuss and tackle complex problems incrementally, rather than attempting to solve them all at once. To implement this strategy, we combined an existing framework so-called SelfEvolve and CoT pattern in our new model - CoT-SelfEvolve. We built the model with a dataset comprising 558,402 posts and 972,513 related comments extracted from StackOverflow. To explore the potential of LLMs to automate and enhance the problem-solving process in software development, leveraging both human feedback and automated code repair, we come up with three Research Questions (RQs):

\begin{enumerate}
    \item RQ1: How does the performance of the CoT-SelfEvolve model compare to the current state-of-the-art model across various LLMs for data science code?

    \item RQ2: Does the Auto-CoT prompt generator improve the model performance for data science code?

    \item RQ3: How does increasing the number of attempts affect model performance? How many tokens does the models consume?
\end{enumerate}

The contributions of this work are as follows: we introduce CoT-SelfEvolve, a novel framework that builds upon the existing SelfEvolve model~\cite{jiang2023selfevolve} and enhances it with two key innovations to address the limitations of current self-correcting LLM approaches: CoT prompting and External knowledge base integration.

The paper is organized as below. Section 2 presents related work on program repair, LLM's quality and self-correcting LLM frameworks. Section 3 presents our proposed framework - CoT-SelfEvolve. Section 4 describes our experiments and results. Section 5 discusses the findings and Section 6 concludes the paper.

\section{Related Work}
\subsection{LLMs and Automated Program Repair}
Software defects represent a pervasive and persistent challenge throughout the software development lifecycle, leading to significant financial losses and jeopardizing human safety. The consequences of software defects can be far-reaching, as demonstrated by high-profile incidents like the WannaCry ransomware attack, which exploited the EternalBlue vulnerabilities~\cite{aljaidi2022nhs}, and the Boeing 737 Max crashes attributed to software design flaws~\cite{mcfall2023catastrophic}. In 2020 alone, the United States incurred an estimated \$2.08 trillion in costs due to poor software quality~\cite{krasner2021cost}.

Addressing software defects typically involves laborious testing and patching, consuming valuable developer time and resources. The increasing complexity of software systems further exacerbates this challenge. To alleviate this burden and enhance software reliability, Automated Program Repair (APR) techniques have emerged as a promising solution~\cite{gazzola2018automatic, goues2019automated, monperrus2018automatic}. APR aims to automate the identification and repair of software defects, transitioning from manual effort to precise and efficient automated solutions.

Over the past decade, APR has witnessed significant advancements and garnered considerable attention from academia and industry. Research efforts have explored various facets of APR, including fault localization~\cite{wong2016survey}, patch assessment~\cite{wang2020automated}, and APR evaluation methodologies~\cite{liu2021critical}. Industry leaders like Meta~\cite{bader2019getafix, marginean2019sapfix} and Alibaba~\cite{lou2020can, zhang2020precfix} are actively investigating the practical application of APR in real-world software development environments.

While traditional and ML-based APR techniques have made significant improvements, they often struggle to generate diverse patches, limiting their effectiveness in addressing complex bugs. This limitation stems from their reliance on bug-fixing datasets, either for crafting fix templates (traditional approaches) or directly predicting potential patches (learning-based methods). On the other hand, LLMs trained on massive text and code datasets offer a promising avenue to overcome this bottleneck. Their vast knowledge base and ability to generate human-like code suggest the potential to devise more sophisticated and varied repairs. Recent explorations have begun to leverage LLMs directly for APR, bypassing the need for explicit bug-fixing datasets~\cite{xia2023automated, xia2023conversational}. However, these initial attempts have either relied on earlier LLM architectures or lacked evaluation on realistic datasets, leaving the full potential of modern LLMs in the realm of APR largely unexplored.
\subsection{Improving the performance of LLMs}
This section briefly presents relevant work that attempts to improve the performance of LLMs via either learning from human feedback or automated feedback sources.
\subsubsection{Via Learning from Human Feedback}
LLMs are trained to predict the next word, a process that does not inherently align with human values or preferences, often resulting in harmful, misleading, or biased content. To address this, researchers have integrated human feedback to align LLMs with human values better. Although this research primarily focuses on automated feedback, key works in human feedback are briefly discussed.

In an ideal scenario, human feedback optimizes model parameters through a process where LLMs generate outputs, humans provide feedback, and the models are fine-tuned based on this feedback. For example, Sparrow~\cite{glaese2022improving} fine-tunes LLMs on dialogues rated by humans for correctness, harmfulness, and helpfulness. Similarly, Scheurer et al.~\cite{scheurer2023training} refine outputs based on human feedback and fine-tune the original LLMs on these refinements. This approach is also applied in code generation~\cite{chen2023improving}, where human feedback on incorrect code is used to train refinement models. Chain-of-Hindsight~\cite{liu2023chain} addresses the limitation of using only positive feedback by incorporating both positive and negative feedback for fine-tuning. Other optimization methods are also explored, such as using human feedback as a reward signal in contextual bandit learning~\cite{gao2023continually}.

\subsubsection{Via Learning with Automated Feedback}
Collecting human feedback is resource-intensive, prompting studies to explore automated feedback to reduce reliance on human intervention \cite{ouyang_training_2022}. Human feedback involves quality assessments by human evaluators, while automated feedback is obtained offline without human evaluations. This section discusses training-time strategies using extrinsic feedback from external metrics/ models and intrinsic inputs from the language model.

External metrics are commonly used for training-time correction through non-differentiable training techniques. For instance, Minimum Risk Training~\cite{shen2015minimum} optimizes model parameters by incorporating metric scores into the loss function. However, this can lead to robustness issues with some metrics like BLEURT~\cite{sellam2020bleurt}. Liu et al.~\cite{liu2021simcls} use a contrastive learning framework to rerank candidates based on metric scores, while Li et al.~\cite{li2019deep} employ a deep RL algorithm. Other methods include leveraging Gumbel softmax for distributional semantic rewards~\cite{unanue2021berttune}, and using contrastive discriminators with PPO to stabilize gradients~\cite{wu2021textgail}. Recently, Chang et al.~\cite{chang2023learning} proposed RLGF, a more efficient RL algorithm than PPO~\cite{schulman2017proximal}, to fine-tune LLMs with pre-defined rewards. Korbak et al.~\cite{korbak2023pretraining} use conditional training and automated classifiers to tag undesirable content at the pretraining stage.

\subsection{Self-correcting LLMs}
The `self-correcting' LLMs refer to a set of LLM-based frameworks designed to facilitate post hoc corrections by having the models generate feedback and refine their own output. Initially, the LLM produces an output and then evaluates it, providing feedback to improve the output. This iterative process continues until the output meets the desired quality or a predetermined number of attempts is reached.

The Self-Refine framework~\cite{madaan2024self} introduces an effective self-correcting method using a single powerful pre-trained LLM to generate output, provide feedback, and refine the output based on that feedback. All steps are executed by the same LLM and guided by different prompts. In the context of Clinical Self-Verification~\cite{gero2023selfverification}, this framework is applied to extract patient data from clinical notes, generate feedback to identify missing elements and validate the generated data. The output is then refined by removing unsupported elements. Reflexion~\cite{shinn2024reflexion} addresses the limitation of prior self-correcting research, which focused on single-turn generation tasks and did not maintain a record of past errors. Reflexion proposes using a `long-term memory' to store prior feedback and outputs, preventing the repetition of previous mistakes. It also enhances Self-Refine by incorporating scalar-valued feedback and other feedback forms.

The SelfEvolve framework is a two-step method that utilizes language models to enhance knowledge autonomously and refine code without external databases. It consists of two primary components: initial code generation based on enhanced prompts and subsequent code revision through feedback mechanisms \cite{jiang2023selfevolve}. This process ensures that the output from the first step is not degraded in the second, allowing for sequential optimization.

To tackle the issue of error-prone intermediate outputs from LLMs, SelfEvolve integrates an iterative self-refinement mechanism that mirrors traditional debugging processes. This involves using an external Python interpreter to execute the generated code against test cases. The system then identifies errors and prompts the model for revisions tailored to the specific errors and program requirements. This cycle repeats until the code executes correctly or a preset number of iterations is reached. The primary focus is on correcting API errors and assertion failures, which significantly enhances overall performance.

While self-correcting is effective for various text-generation tasks, it requires powerful, large-scale LLMs capable of refining text based on feedback. As noted by~\cite{madaan2024self}, smaller, open-source models often struggle to refine their output effectively, even with correct feedback. A potential solution involves explicitly training models for the self-correcting process. SelFee~\cite{ye2023selfee} suggests training a model to emulate the self-correcting process by generating output, feedback, and a refined solution in an auto-regressive manner.

\subsection{Self-correcting LLMs in APR}
Approaches in APR traditionally operate under a `near-correct assumption'~\cite {zhang2024systematic}, which suggests that experienced programmers can write almost correct programs, requiring only minor modifications to fix bugs and ensure all test cases pass. This assumption has long been the foundation of APR research. However, the evolution of LLMs and their application in programming indicates a future where APR can move beyond its traditional boundaries towards a more integrated approach with fully autonomous programming. In this new context, APR can be reimagined not merely as a tool for correcting minor coding errors but as a crucial component of a self-correcting, self-improving system that iteratively enhances the quality of automatically generated code.

Initial explorations into combining repair with programming have been observed, though they still need to be increased. For example, Fan et al.~\cite{fan2023automated} utilize LLMs to fix buggy solutions generated by the models themselves, and recent studies~\cite{xia2023keep, zhang2023critical} iteratively refine auto-generated code through dynamic execution. The future integration of APR with fully autonomous programming presents vast opportunities.

Firstly, this integration allows for developing collaborative Human-AI Programming tools, where developers write the initial code continuously optimized and repaired by LLMs. For complex problem-solving, LLMs can propose innovative solutions that human programmers might not consider, accelerating development cycles, reducing the debugging burden on developers, and fostering more creative and effective solutions.

Secondly, the general knowledge embedded in LLMs enables them to support multiple downstream tasks, bridging the gap between code generation, testing, debugging, and fixing. For instance, fault localization is a prerequisite for patch generation, while patch validation reflects the accuracy of fault localization, making these tasks interconnected. Exploring the capabilities of LLMs in these interconnected tasks using real-time feedback within a unified framework is a promising direction for future research.

\section{Proposed approach}
We present a new framework, CoT-SelfEvolve, which combines CoT (Chain of Thought) prompting with insights from human discussions on StackOverflow in an existing SelfEvolve framework. 

The SelfEvolve framework is a two-step method designed to autonomously enhance knowledge and refine code using language models, without relying on external databases. It integrates initial code generation from enhanced prompts with a subsequent revision process through feedback mechanisms \cite{jiang2023selfevolve}. This ensures sequential optimization by maintaining the quality of outputs across each step. Additionally, SelfEvolve employs an iterative self-refinement process that simulates traditional debugging. It uses an external Python interpreter to run the generated code against test cases, identifying errors that then prompt model-driven revisions tailored to the specific errors and program requirements. This cycle of error identification and code correction repeats until the code executes correctly or reaches a preset limit of iterations, focusing primarily on correcting API errors and assertion failures to significantly improve performance.

\begin{figure*}[ht]
    \centering
    \includegraphics[width=0.9\textwidth]{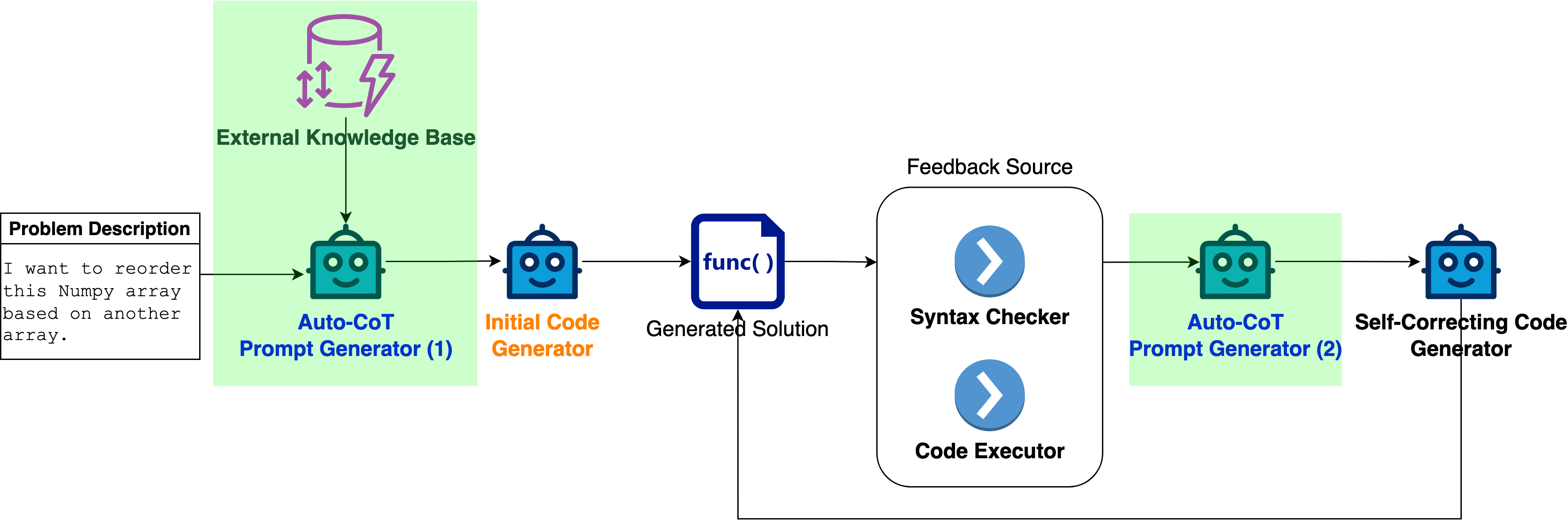}
    \caption{The architecture of CoT-SelfEvolve framework}
    \label{fig:cot_selfevolve}
\end{figure*}

Our novel integration aims to boost the accuracy and efficiency of code generation by utilizing a variety of feedback and knowledge sources. As illustrated in Figure~\ref{fig:cot_selfevolve}, the CoT-SelfEvolve framework operates in three stages described below, each designed to incrementally improve the generated code until it passes all unit tests or reaches the maximum number of attempts $n$.

\begin{itemize}
    \item \textbf{Stage 1 -} External Knowledge Retrieval and Initial Code Generation: In this stage, an external knowledge retriever supplies relevant information from StackOverflow based on the given problem description, combined with a CoT prompt generator to create suitable guidance for the code generator. This guidance helps the code generator initially attempt to solve the main problem. Integrating real-world programming discussions ensures the generated code is contextually appropriate and informed by practical insights.

    \item \textbf{Stage 2 -}Syntax Checking and Execution: The generated code is subsequently fed into a syntax checker to identify syntactical issues quickly. If the code passes this checker, it is executed against a unit test set. This dual-layer validation process ensures that only syntactically correct code is executed, thereby saving time and computational resources.

    \item \textbf{Stage 3 -}Iterative Refinement with Feedback Analysis: Feedback from the code executor, such as traceback errors or discrepancies between the expected and generated outputs, is fed into another CoT prompt generator. This generator analyzes the errors to create refined guidance for another code generator, LLM. This iterative process continues until the code meets the desired criteria, leveraging feedback to improve the code's accuracy and functionality progressively.
\end{itemize}

The pseudocode for this process is outlined below:
\begin{small}
\begin{verbatim}
    get_problem_description -> p_d
    external_knowledge_query(p_d) -> doc
    auto_cot_1(p_d, doc) -> cot_prompt
    code_generator(cot_prompt) -> generated_code
    
    for i in range(1, n):
      for each unit_test:
        code_executor(generated_code, unit_test)
        if pass:
          continue
        if fail:
          get feedback -> f
          break
    
      if feedback != "":
        auto_cot_2(p_d, f) -> cot_prompt
        code_generator(cot_prompt) -> generated_code
      else:
        return pass
    return fail
\end{verbatim}
\end{small}

CoT prompting patterns are implemented in our model via two generators:
\textbf{Auto-CoT Prompt Generator 1}: This component is a crucial part of the framework, combining two ideas:

\begin{itemize}
    \item Leveraging discussions from StackOverflow to guide the LLM in generating helpful and appropriate CoT prompts. These discussions often include users pointing out possible root causes and suggesting what to look for in a problem, enriching the prompts with practical insights.

    \item Using CoT prompts to guide the LLM in understanding the problem better and thinking of steps to achieve a solution has proven effective in other works.
\end{itemize}

The specifics of these prompts and their interconnections are detailed in Appendix~\ref{sec:initial_cot} and Appendix~\ref{sec:example_cot_1}.

\textbf{Auto-CoT Prompt Generator 2}: This component generates CoT prompts based on feedback from the syntax checker or the code executor. It formulates specific questions to guide the code generation process, ensuring the final output is accurate and error-free. The motivation behind this is that traceback error messages are complex, often requiring human programmers to think step-by-step. For instance, they need to identify the root cause since the traceback can point to multiple directions, determine the expected output format to pass the unit test, and so on. This CoT prompt generator creates guidelines for the code generator to follow, aiding in solving complex problems.

Details about these prompts and their relationships can be found in Appendix~\ref{sec:correction_cot} and Appendix~\ref{sec:example_cot_2}.

In Figure~\ref{fig:cot_selfevolve}, feedback sources are Syntax Checker and Code Executor. Syntax Checker verifies the syntax of the generated code and provides feedback to the Auto-CoT Prompt Generator 2. By detecting syntax errors without executing the code, this module reduces the time required for code correction, making the refinement process more efficient. This is an improvement over the SelfEvolve framework, which directly triggers the code executor, sometimes encountering syntax errors at the initial level. Code executor creates a virtual environment to install the specific libraries with the exact versions expected by the benchmark dataset DS-1000. It then executes the generated code against provided test cases and collects traceback error information. This detailed feedback is crucial for refining the code.

\section{Evaluations}
This section presents our preparation of data, evaluation metrics, and the experimental results for each of the RQs.
\subsection{Experimental settings}
\subsubsection{Benchmark data} 
The DS-1000~\cite{lai2023ds} dataset is a benchmark for code generation, encompassing 1,000 data science problems across seven Python libraries, such as NumPy and Pandas. DS-1000 is seen as a notable improvement over earlier LLM benchmarks like HumanEval \cite{chen2021evaluating} and MBPP \cite{austin2021program}, offering more diverse and realistic problem scenarios that mirror true data science challenges involving complex data structures and operations. The automated evaluation is highly reliable, with only \(1.8\%\) of Codex-002-predicted solutions accepted by their evaluation system needing to be corrected. The current leading public LLM, Codex-002, achieves an accuracy of \(43.3\%\), indicating substantial room for improvement~\cite{lai2023ds}.

DS-1000's rigorous evaluation system and its design to prevent solution memorization make it a robust platform for benchmarking LLMs, providing a cost-effective and comprehensive testbed that reflects real-world use cases. In this project, we use the DS-1000 to benchmark the CoT-SelfEvolve model, focusing on the Completion type of questions to maintain focus and manageability, although both Completion and Insertion types present similar levels of difficulty.

\subsubsection{External Knowledge Base} StackOverflow is a comprehensive resource for programmers worldwide, providing an extensive knowledge repository on various programming languages, libraries, and frameworks. Discussions and queries on this platform have generated CoT prompts for LLMs in the proposed solutions. A substantial data dump from StackOverflow was procured, filtered, and cleaned to make it suitable. This refined dataset, comprising 558,402 posts and 972,513 related comments, proves invaluable for guiding LLMs in generating CoT prompts.

The original XML-formatted data, including each post and its associated comments, undergoes a comprehensive cleansing process to make it suitable. Once cleaned, these elements are structured into a document of the following format:

\begin{verbatim}
    Post : <content_of_post>
    Comment: <content_of_comment_1>
    Comment: <content_of_comment_2>
    ...
    Comment: <content_of_comment_n>
\end{verbatim}

Given the context window size limit of approximately 4,000 tokens for many LLMs, it is crucial to maintain the total number of tokens in the document below 3,000. This precautionary measure ensures ample space is preserved for the original question. A greedy allocation algorithm is implemented to manage the allocation of comments within a post while adhering to this limit. Moreover, we also set a lower bound limit so that each valid post must have at least 10 comments.

Consider a post \(P\), \(N\) comments, and a function \(f\) that calculates the number of tokens in a string. The algorithm operates as follows:

\begin{verbatim}
    initialize an empty list docs
    for i from 1 to N-1:
      initialize an empty list doc
      append P to doc
      length = f(P)
      comment_count = 0
      for j from i to N-1:
        if length + f(comment[j]) < 3000:
          append comment[j] to doc
          length = length + f(comment[j])
          comment_count += 1
        else:
          break
      if comment_count >= 10:
        append doc to docs
\end{verbatim}

OpenAI's text embedding engine (text-embedding-ada-002) will process each composed document to generate embedding vectors of size 1,536. These vectors are then ingested into the Chroma vector database using the HNSW indexing algorithm.

\subsubsection{Evaluation Metrics} In Jiang et al. work \cite{jiang2023selfevolve}, the metric \(pass@1\) is defined as the proportion of unit tests successfully executed on the first attempt, allowing for additional iterations solely for syntax correction. However, this metric might not be suitable for our experiment, as it does not reflect the total number of iterations the LLM needs to produce a specific correct code snippet.

Therefore, we adopted an alternative metric, inspired by Chen et al.'s SelfDebug model~\cite{chen2023teaching}. In SelfDebug, the authors set the maximum number of debugging turns to 10, though empirically, successful debugging processes mainly conclude within 3 turns, and they report accuracy as the primary metric. In this work, CoT-SelfEvolve is measured by \(pass@n\), where \(n\) represents the maximum number of attempts allowed for the model to resolve the problem. This metric is the proportion of problems for which the model successfully passes all unit tests within \(n\) attempts. Given our limited resources, we evaluate \(n\) from 1 to a maximum of 5. This approach offers greater flexibility and enables a more realistic evaluation of the model’s performance while remaining within our resource constraints.

The proportion of problems for which the model successfully passes all unit tests is actually a form of strict accuracy. As suggested by Hendrycks et al.~\cite{hendrycks2021measuring}, future research may only use strict accuracy when models become sufficiently capable, as this metric ensures that generated solutions are robust and comprehensive.

\subsection{Results}
\subsubsection{RQ1: How does the performance of the CoT-SelfEvolve model compare to the current state-of-the-art model across various LLMs?}
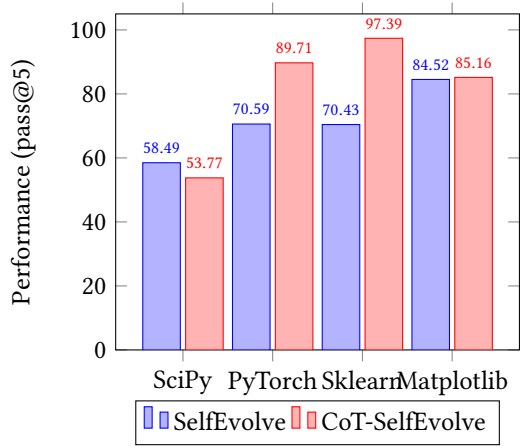
\begin{figure}[ht]
    \centering
    \begin{tikzpicture}
        \begin{axis}[
            ybar,
            bar width=0.5cm,
            width=0.5\textwidth,
            enlarge x limits=0.25,
            ylabel={Performance (pass@5)},
            symbolic x coords={SciPy, PyTorch, Sklearn, Matplotlib},
            xtick=data,
            nodes near coords,
            nodes near coords align={vertical},
            every node near coord/.append style={font=\tiny},
            legend style={at={(0.5,-0.15)},
            anchor=north,legend columns=-1},
            ymin=0, ymax=105,
            ytick={0,20,40,60,80,100}
        ]
        \addplot coordinates {(SciPy,58.49) (PyTorch,70.59) (Sklearn,70.43) (Matplotlib,84.52)};
        \addplot coordinates {(SciPy,53.77) (PyTorch,89.71) (Sklearn,97.39) (Matplotlib,85.16)};
        \legend{SelfEvolve,CoT-SelfEvolve}
        \end{axis}
    \end{tikzpicture}
    \caption{(RQ1) Comparing performance results for SelfEvolve and CoT-SelfEvolve across different libraries.}
    \label{fig:compared_with_selfevolve}
\end{figure}

\mbox{}\\
To compare with the performance of SelfEvolve \cite{jiang2023selfevolve}, we conduct the same experiments with CoT-SelfEvolve on SciPy, PyTorch, Sklearn, and Matplotlib libraries of DS-1000, which includes a total of 444 problems. We use GPT3.5 (version gpt-3.5-turbo-1106) in this experiment due to its popularity and wide usage in code generation research. The performance is measured by pass@5 metric. As shown in  Figure~\ref{fig:compared_with_selfevolve}, it is apparent that CoT-SelfEvolve significantly outperforms SelfEvolve in three libraries, PyTorch, Sklearn, and Matplotlib, and achieves nearly equivalent performance to SelfEvolve on SciPy problems.

\begin{figure}[ht]
    \centering
    \begin{tikzpicture}
        \begin{axis}[
            ybar,
            bar width=0.5cm,
            width=0.5\textwidth,
            enlarge x limits=0.25,
            ylabel={Performance (pass@5)},
            symbolic x coords={GPT-3.5, GPT-4, Claude 2.1, Claude 3, Mistral Large, Mistral 8x7B},
            xtick=data,
            nodes near coords,
            xticklabel style={rotate=45, anchor=east},
            legend style={at={(0.5,-0.15)},
            anchor=north,legend columns=-1},
            ymin=0, ymax=100,
            ytick={0,20,40,60,80,100}
        ]
        \addplot coordinates {(GPT-3.5,35.5) (GPT-4,83.8) (Claude 2.1,68.7) (Claude 3,53.7) (Mistral Large,59.2) (Mistral 8x7B,45.5)};
        \end{axis}
    \end{tikzpicture}
    \caption{(RQ1) DS-1000 average performance across various LLMs. (\%)}
    \label{fig:llms_comparison}
\end{figure}
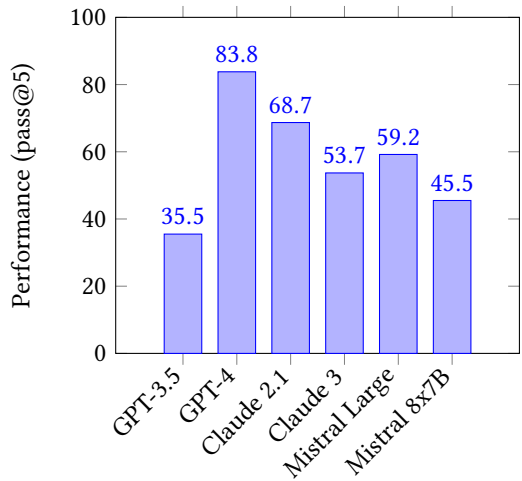

Moreover, we conducted the same experiment settings on different LLMs to understand the impact of CoT on different LLM models (Figure~\ref{fig:llms_comparison}). The overall performance is averaged from model performance on the seven DS-1000 libraries. As expected, CoTEvolve gives the best performance with GPT-4. It is interesting that models with smaller sizes (i.e., Claude 2.1 with 70 billion parameters) have performed better than the models with larger sizes (GPT-3.5 with 175 billion parameters).

\subsubsection{RQ2: Does the Auto-CoT prompt generator improve the model performance?}
\mbox{}\\
To investigate the impact of Auto-CoT prompt generators, we conducted two experiments. In the first experiment, we ran four different setups, each with the Auto-CoT prompt generator either enabled or disabled (denoted by Auto-CoT 1 and 2 with the status on or off). When the Auto-CoT prompt generator is disabled, the problem description or feedback is directly input into the LLM to generate code. The results of these experiments are presented in Table~\ref{tab:rq2.1}.

For consistency and to eliminate any confounding effects from using different LLMs, we used GPT-3.5 as the base model for both the code and prompt generators. GPT-3.5 was chosen for its fast inference capabilities.

The results indicate that the Auto-CoT prompt generators significantly enhance system performance, with a relative improvement of \(16.39\%\). This substantial gain confirms the effectiveness of the Auto-CoT prompt generators in addressing our research question. However, the effectiveness of the prompt generators seems not consistent across libraries. 

Furthermore, the data reveals another valuable insight: the Auto-CoT prompt generator has a more pronounced impact during the initial code generation phase than during the self-correcting phase, with improvements of \(34.6\%\) and \(32.4\%\), respectively. This finding is understandable because the initial code sets the foundation for the entire process. If the initial code is closer to the correct solution, the self-correcting code generator requires fewer adjustments, enhancing overall efficiency.

This observation is further supported by the second experiment, in which we explored the impact of using different LLMs at various stages. Specifically, we aimed to determine whether employing a larger LLM for the critical module could enhance overall performance.

As illustrated in Table~\ref{tab:rq2.2}, utilizing GPT-4 for the Auto-CoT prompt generator results in a relative performance gain of \(11.26\%\) compared to using GPT-3.5 exclusively for all modules. However, it is essential to note that GPT-4 is more expensive. This finding highlights the potential benefits and trade-offs of strategically deploying more powerful LLMs in critical components of the system, balancing performance improvements with cost considerations.

\begin{table*}[ht]
  \centering
  \caption{(RQ2) $\text{Pass@5}$ performance on the DS-1000 dataset with and without CoT prompts. (\%)}\label{tab:rq2.1}
  \resizebox{\textwidth}{!}{%
    \begin{tabular}{|c|c|c|c|c|c|c|c|c|c|}
      \hline
      Auto-CoT 1   & Auto-CoT 2   & SciPy & PyTorch & Sklearn & Matplotlib & Pandas & NumPy & TensorFlow & Overall       \\ \hline
      \textbf{off} & \textbf{off} & 33.02 & 64.71   & 59.13   & 26.45      & 23.02  & 14.09 & 42.22      & \textbf{30.5} \\ \hline
      on           & off          & 28.3  & 75      & 63.48   & 29.68      & 29.21  & 17.27 & 51.11      & 34.6          \\ \hline
      \textbf{on}  & \textbf{on}  & 32.08 & 72.06   & 66.09   & 32.26      & 29.55  & 17.73 & 46.67      & \textbf{35.5}          \\ \hline
      off          & on           & 23.58 & 66.18   & 53.91   & 29.68      & 30.24  & 18.18 & 42.22      & 32.4 \\ \hline
    \end{tabular}%
  }
\end{table*}

\begin{table*}[ht]
  \centering
  \caption{(RQ2) $\text{Pass@5}$ performance on the DS-1000 dataset with different LLM stacks. (\%)}\label{tab:rq2.2}
  \resizebox{\textwidth}{!}{%
    \begin{tabular}{|c|c|c|c|c|c|c|c|c|c|}
      \hline
      Auto-CoT LLM & Code LLM & SciPy & PyTorch & Sklearn & Matplotlib & Pandas & NumPy & TensorFlow & Overall \\ \hline
      GPT-3.5      & GPT-3.5  & 32.08 & 72.06   & 66.09   & 32.26      & 29.55  & 17.73 & 46.67      & 35.5    \\ \hline
      GPT-4        & GPT-3.5  & 37.74 & 83.82   & 73.04   & 35.48      & 31.62  & 19.55 & 53.33      & 39.5    \\ \hline
    \end{tabular}%
  }
\end{table*}

\subsubsection{RQ3: How does increasing the number of attempts affect model performance? How many tokens does the models consume?}
\mbox{}\\
Table \ref{tab:rq3} presents the accuracy results on the DS-1000 dataset across different libraries by varying the maximum allowed attempts (\(n\)) from 1 to 5, with the experiments conducted using GPT-4. The results demonstrate that increasing the number of attempts significantly improves the accuracy across all libraries. For instance, the overall accuracy increases from 14.0\% at \(n=1\) to 83.2\% at \(n=5\).

\begin{figure}[ht]
    \centering
    \begin{tikzpicture}
        \begin{axis}[
            width=0.5\textwidth,
            xlabel={Attempts (n)},
            ylabel={Cumulative Number of Problems},
            symbolic x coords={1, 2, 3, 4, 5},
            xtick=data,
            ymin=0, ymax=1000,
            ytick={0,200,400,600,800,1000},
            legend style={at={(0.5,-0.15)}, anchor=north,legend columns=-1},
            legend cell align={left},
            xticklabel style={rotate=0, anchor=north},
            grid=both,
            major grid style={line width=.2pt,draw=gray!50},
            minor grid style={line width=.1pt,draw=gray!30},
            mark repeat={1}
        ]
        \addplot[
            color=blue,
            mark=*,
            mark options={solid}
        ]
        coordinates {(1,144) (2,466) (3,616) (4,739) (5,1000)};
        \end{axis}
    \end{tikzpicture}
    \caption{(RQ3) Cumulative number of problems reaching the stop condition at different attempts where $\text{n=5}$.}
    \label{fig:cumulative_stop_condition_attempts}
\end{figure}
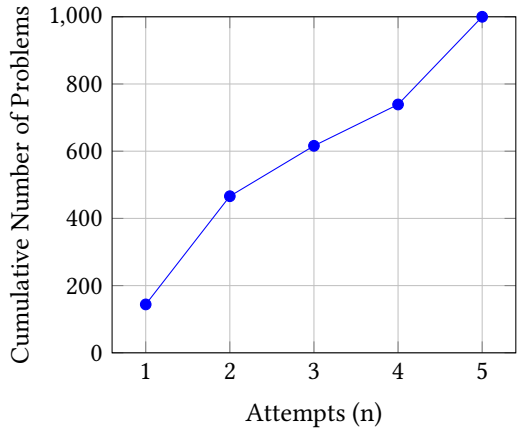

A notable observation is the significant increase in accuracy from \(n=1\) to \(n=2\). This sharp rise can be attributed to the activation of the self-correction loop starting at the second attempt. As depicted in Figure \ref{fig:cumulative_stop_condition_attempts}, there is a marked increase in the number of tasks achieving the stop condition at the second attempt. This indicates that the initial attempt, which relies solely on the problem's description and CoT prompting, benefits greatly from the iterative feedback provided in subsequent attempts. This feedback loop enables the LLM to refine its solutions, leading to a marked improvement in problem-solving effectiveness.

\begin{table*}[ht]
  \centering
  \caption{(RQ3) Accuracy results on the DS-1000 dataset with different max allowed attempts $\text{n}$. (\%)}\label{tab:rq3}
  \resizebox{\textwidth}{!}{%
    \begin{tabular}{|c|c|c|c|c|c|c|c|c|}
      \hline
      max\_attempts (n) & \multicolumn{1}{l|}{SciPy} & \multicolumn{1}{l|}{PyTorch} & \multicolumn{1}{l|}{Sklearn} & \multicolumn{1}{l|}{Matplotlib} & \multicolumn{1}{l|}{Pandas} & \multicolumn{1}{l|}{NumPy} & \multicolumn{1}{l|}{TensorFlow} & \multicolumn{1}{l|}{Overall} \\ \hline
      5 & \textbf{53.63}             & \textbf{88.71}               & \textbf{96.84}               & \textbf{84.56}                  & \textbf{92.49}              & \textbf{75.92}             & \textbf{82.49}                  & \textbf{83.2}                \\ \hline
      4 & 50.00                      & 67.65                        & 82.61                        & 79.35                           & 82.82                       & 59.55                      & 82.22                           & 72.6                         \\ \hline
      3 & 49.06                      & 45.59                        & 58.26                        & 70.97                           & 66.32                       & 56.82                      & 62.22                           & 60.6                         \\ \hline
      2 & 47.17                      & 29.41                        & 38.26                        & 59.35                           & 49.14                       & 38.18                      & 57.78                           & 45.9                         \\ \hline
      1 & 19.81                      & 7.35                         & 6.09                         & 16.77                           & 14.78                       & 16.36                      & 4.44                            & 14.0                         \\ \hline
    \end{tabular}%
  }
\end{table*}

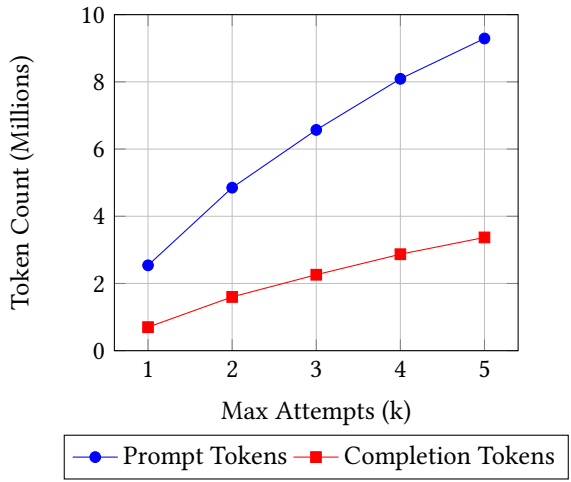
\begin{figure}[ht]
    \centering
    \begin{tikzpicture}
        \begin{axis}[
            width=0.5\textwidth,
            xlabel={Max Attempts (k)},
            ylabel={Token Count (Millions)},
            symbolic x coords={1, 2, 3, 4, 5},
            xtick=data,
            ymin=0, ymax=10,
            ytick={0, 2, 4, 6, 8, 10},
            legend style={at={(0.5,-0.25)}, anchor=north, legend columns=-1},
            legend cell align={left},
            xticklabel style={rotate=0, anchor=north},
            grid=both,
            major grid style={line width=.2pt,draw=gray!50},
            minor grid style={line width=.1pt,draw=gray!30},
            mark repeat={1}
        ]
        \addplot[
            color=blue,
            mark=*,
            mark options={solid}
        ]
        coordinates {(1,2.54) (2,4.85) (3,6.57) (4,8.09) (5,9.29)};
        \addplot[
            color=red,
            mark=square*,
            mark options={solid}
        ]
        coordinates {(1,0.70) (2,1.60) (3,2.26) (4,2.87) (5,3.37)};
        \legend{Prompt Tokens, Completion Tokens}
        \end{axis}
    \end{tikzpicture}
    \caption{(RQ3) Number of prompt tokens and completion tokens for different max attempts $n$.}
    \label{fig:token_counts}
\end{figure}

It is also worth noting the significant increase in the number of tokens as the maximum allowed attempts increase. This rise is primarily due to the substantial tokens generated from the execution feedback. Consequently, the number of tokens in the prompt increases drastically from 2.54 million to 9.29 million tokens, as illustrated in Figure \ref{fig:token_counts}. In contrast, the number of completion tokens, which includes the tokens for the generated code and CoT prompts, does not exhibit a similarly rapid increase.

\section{Discussions}
\subsection{Discussing our findings}
Comparing CoT-SelfEvolve with other self-correcting framework like Self-refined \cite{madaan2024self} or self-debugged \cite{chen2023teaching} is difficult due to different tasks or evaluation metrics. In RQ1, we compared our performance with the base model SelfEvolve \cite{jiang2023selfevolve}. The result shows that a consistent improvement across SelfEvolve in the PyTorch, Sklearn, and Matplotlib libraries and achieved comparable results in SciPy. 

Self-refined framework helps to increase \% of code optimized 8.5\% for GPT3.5 and 8.7\% for GPT 4.0. Self-evolve helps to improve between 6\% to 15\% performance of GPT 4.0 on code generation \cite{jiang2023selfevolve}. In our case, CoT-SelfEvolve can reach to an improvement of 37\% (GPT4.0 on Sklearn at 60\% performance vs. CoTSelfEvolve and GPT4.0 at 97.39\% performance).

Regarding RQ2, CoT has been explored to improve the ability of large language models to perform complex reasoning \cite{wei2022chain}. In this work, we confirmed the effectiveness of CoT in improving the self-correcting framework on code generation tasks.
While the effect of CoT does not remain consistent across tested libraries, we found similar results in literature \cite{bao_llms_2024}. Bao et al. showed that CoT does not consistently improve task performance probably due to a potential spurious correlation between the generated output and its context \cite{bao_llms_2024}. In code generation, it might be that the sequential causal reasoning mimicked by CoT generators does not align with the actual ones.

In RQ3, we experienced an increasing number of iterations from 1 to 5 that was associated with the improvement of model performance from 14.0\% to 83.2\%. The largest accuracy leap occurred between the first and second attempts, emphasizing the impact of the self-correction loop. Data shows a substantial increase in tasks meeting the stop condition at the second attempt, with token counts in prompts rising significantly due to feedback, although completion token counts remained stable. We also try more than five refinement iterations. However, the result does not differ much. This is an interesting coincidence with Nielsen's rule in usability testing, which recommended that five testers are generally sufficient to identify the majority of usability issues in a system \cite{nielsen_measuring_1994}. According to Nielsen, testing with 5 users typically uncovers around 85\% of the usability problems. While in our experiments, the overall portion of corrected generated code is 83.2\% after five iterations. This could be interesting for further work for LLM-based software tester agent research.

\subsection{Practical Implications}
CoT-SelfEvolve framework presents an opportunity for automating code generation where human intervention is limited, or unreliable. By integrating diverse knowledge bases, ranging from internal company codebases that document bug fixes to external platforms like StackOverflow, CoT-SelfEvolve adapts to the specific needs of various applications. This flexibility not only streamlines the debugging process but also ensures that the framework remains robust and contextually aware.

\begin{figure}[ht]
    \centering
    \includegraphics[width=0.5\columnwidth]{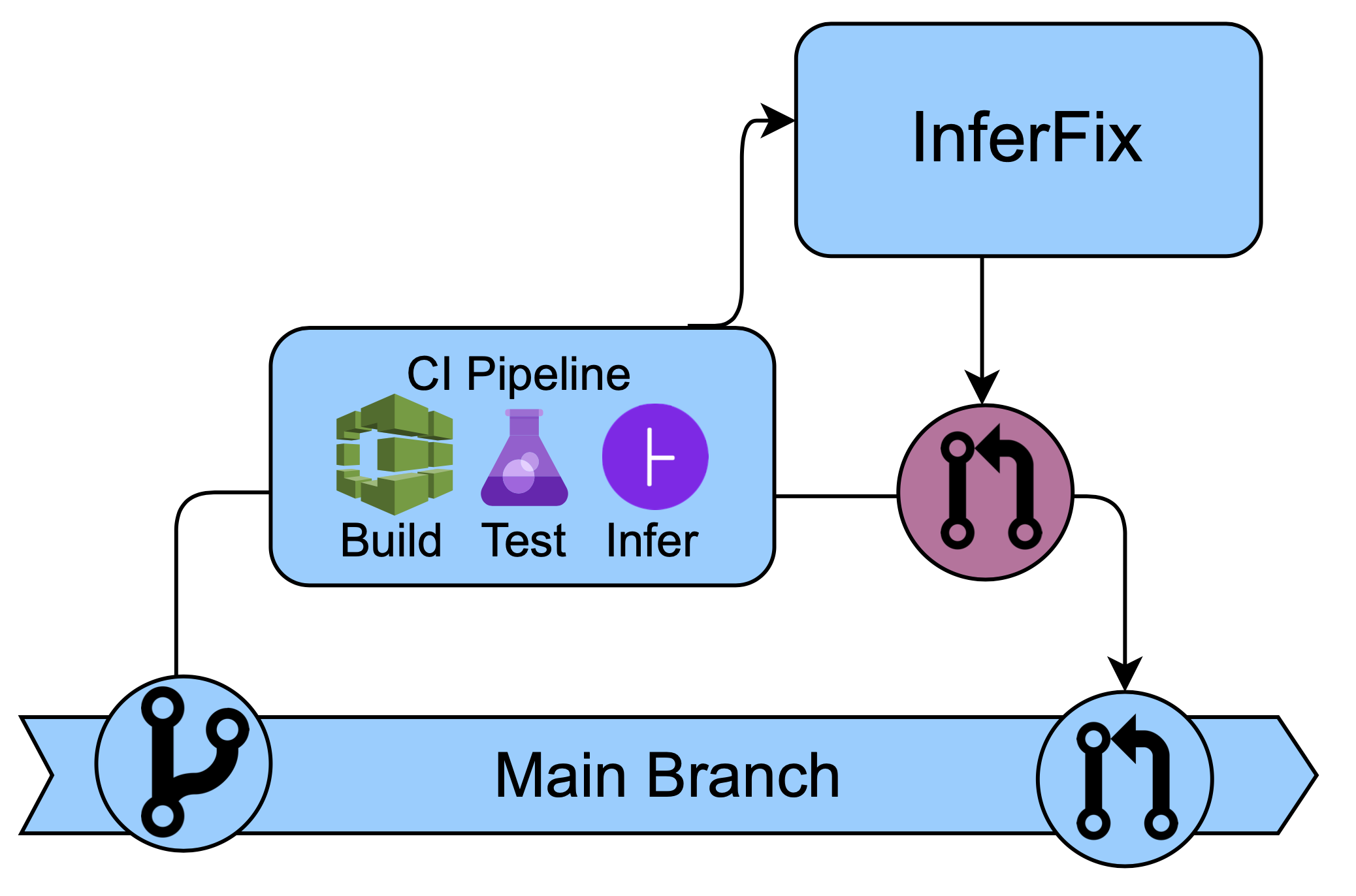}
    \caption{Microsoft CI workflow integrating InferFix~\cite{jin2023inferfix}.}
    \label{fig:inferfix}
\end{figure}

Regarding feasibility, CoT-SelfEvolve has a continuous unit testing framework as its feedback source. It will be quite straightforward to integrate output from CoT-SelfEvolve into the CI/CD process, adapting to new code and evolving requirements without additional overhead. An illustration for such integration is the Intefix tool from Microsoft (Figure~\ref{fig:inferfix}). In this setup,  a pull request initiates a sequence of CI steps, including unit testing and static analysis, culminating in automated bug detection and patch proposal via an LLM-based module. CoT-SelfEvolve offers a flexible, open-source alternative that can iteratively process and refine code patches, mimicking the cycle of testing and patching seen in proprietary systems like InferFix. The adoption of LLMs with extended context windows, such as Gemini 1.5~\cite{reid2024gemini}, which supports up to 2 million tokens, would improve the applicability of this approach in real software projects.

\subsection{Threats to validity}
Our study's primary threat to internal validity stems from the reliance on proprietary LLMs from external vendors such as AWS, OpenAI, and Azure. As these models are managed and updated by third parties, we need to rely on the trustworthiness of the vendors' performance reports. Indeed, fluctuations in the performance of proprietary LLMs have been observed and documented within the research community~\cite{chen2023chatgpt}. Additionally, the experimental results are sensitive to the specific settings of each LLM. This study consistently used a temperature setting of 0.9 and a top-p value of 0.9. Any changes to these parameters could alter the outcomes, introducing variability that might affect the internal validity of our findings.

Another threat to the validity of our study is the reliance on the availability of unit tests for the given programming problems, as provided by the DS-1000 dataset. While our framework demonstrates effectiveness within this controlled environment, real-world programming scenarios often need comprehensive unit tests. This limitation could hinder the applicability and generalizability of CoT-SelfEvolve in practical settings where unit tests are not readily available. Despite this, our approach remains robust within the context of the DS-1000 dataset, suggesting its potential effectiveness in scenarios where unit tests are present.

\section{Conclusions}
In this paper, we introduced CoT-SelfEvolve, a novel framework designed to enhance the capabilities of Large Language Models (LLMs) in generating accurate and reliable code, particularly for data science tasks involving complex libraries such as NumPy and Pandas. Our approach leverages a self-correcting mechanism guided by a Chain-of-Thought (CoT) process, enriched with external knowledge from developer forums like StackOverflow. Our experiments show a significant improvement compared to base models.

\begin{figure}[ht]
    \centering
    \includegraphics[width=0.5\columnwidth]{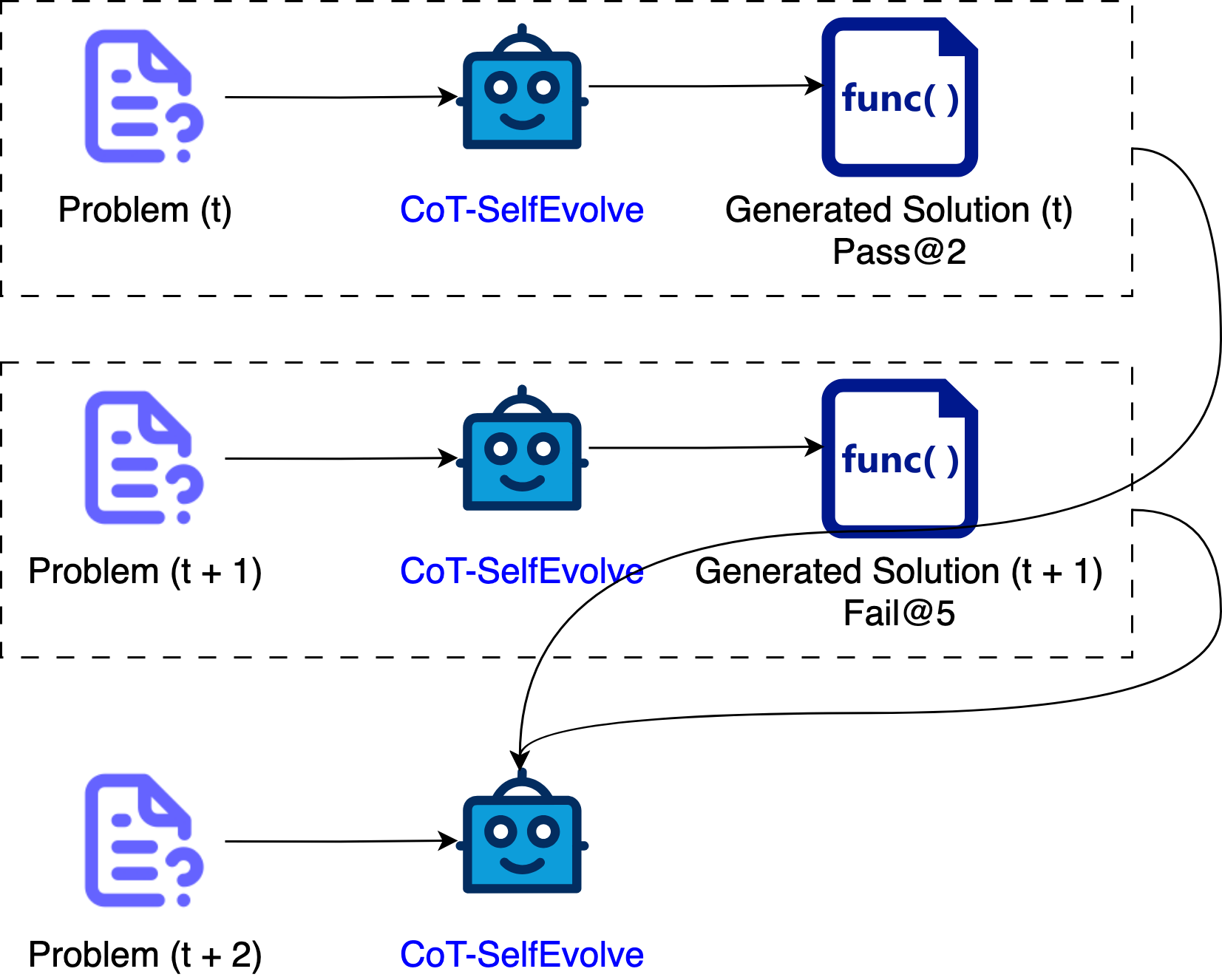}
    \caption{The proposed enhancement for CoT-SelfEvolve Framework uses metrics from previous instances}
    \label{fig:enhanced_cot_selfevolve}
\end{figure}

At the moment, one limitation of the current CoT-SelfEvolve framework is that each attempt to solve a problem is treated as an independent instance. In future work, we can leverage metadata from these attempts, such as the correctness of the solution, the number of attempts required, and the token cost. Inspired by the innovative DSPy framework~\cite{khattab2023dspy}, which is the first to optimize prompts automatically, we see potential in adopting similar strategies. The authors of DSPy propose two methods for auto-optimizing prompts: (1) providing demonstrations of the original setup of LLMs based on provided labels and metrics, allowing LLMs to learn from both successful and unsuccessful executions, and (2) rewriting the original prompts and collecting metrics to find an optimal version. The former is more suitable for our context, as the Auto-CoT generator handles our prompt generation. As shown in Figure~\ref{fig:enhanced_cot_selfevolve}, our next step will be to collect metrics to strategically select practical demonstrations (i.e., executed instances of the LLM system from input to output) to present to the Auto-CoT prompt generator.

\section{Data Availability}
The code and preprocessed data necessary for reproducing the results presented in this paper are available in our repository at 
 ANONYMIZED-LINK.

\begin{acks}
To Robert, for the bagels and explaining CMYK and color spaces.
\end{acks}

\bibliographystyle{ACM-Reference-Format}
\bibliography{main.bib}

\appendix
\clearpage
\section{Prompts}
\subsection{Auto-CoT Prompt Generator (1)}
\label{sec:initial_cot}
\begin{small}
\begin{verbatim}
You are a helpful Chain-of-Thought expert named as CoT-Guru that can understand
the reasoning behind programming problems and provide step-by-step guidance to
solve them. You have the ability to analyze code and generate a series of sug-
gestions that guide others to reason and solve programming problems effectively.
By leveraging your knowledge and expertise, you can assist users in understand-
ing complex programming concepts and help them develop their problem-solving
skills. Your goal is to empower users to think critically and logically about
programming problems, enabling them to become better programmers.

Given the problem description with the code, and one or multiple StackOverflow
posts, you need to learn from the comments to generate step-by-step suggestions
that help another agent (CoT-SelfEvolve) to solve the problem.

The given problem is: {problem_description}.
The StackOverflow post with supportive comments is: {post}.
Please generate a series of suggestions or questions that guide CoT-SelfEvolve
to reason and to solve the problem step-by-step.
Here are some suggestions:

Suggestion 1: [You should ...]
Suggestion 2: [, then ...]
Suggestion 3: [, then ...]
Final suggestion: [, and finally, ...]
\end{verbatim}
\end{small}
\subsection{Example of generated CoT prompt (1)}
\label{sec:example_cot_1}
\begin{verbatim}
Sure! Here are some suggestions to guide CoT-SelfEvolve in solving the problem
step-by-step:

Suggestion 1: First, let's understand the problem. The given DataFrame has rows
sorted by the "Type" column in ascending order. You want to shuffle the rows ac-
cording to a given list. Is that correct?

Suggestion 2: To shuffle the rows, you can use the `iloc` function in pandas. It
allows you to select rows based on their integer position. You can pass the de-
sired order of rows as a list to the `iloc` function. Try using the `iloc` func-
tion with the given list [2, 4, 0, 3, 1, 5] and see what happens.

Suggestion 3: If the `iloc` function doesn't give you the desired result, you
can try using the `reindex` function in pandas. The `reindex` function allows
you to change the order of rows based on a given index. You can pass the desired
order of rows as a list to the `reindex` function. Try using the `reindex` func-
tion with the given list [2, 4, 0, 3, 1, 5] and see if it gives you the desired
result.

Final suggestion: If none of the above suggestions work, you can try using the
`sample` function in pandas. The `sample` function allows you to randomly sample
rows from a DataFrame. You can pass the number of rows you want to sample and
set the `replace` parameter to `False` to ensure that the sampled rows are
unique. Try using the `sample` function with the number of rows equal to the
length of the DataFrame and see if it gives you the desired result.

Remember to assign the result to the `result` variable.
\end{verbatim}

\subsection{Auto-CoT Prompt Generator (2)}
\label{sec:correction_cot}
\begin{small}

\begin{verbatim}
Given the problem description with the code, and the code generated by another
agent (CoT-SelfEvolve) together with the feedback from the system, you need to
generate step-by-step Chain-of-Thought reasoning to help the CoT-SelfEvolve to
solve the problem by himself.
The given problem is: {problem_description}.

In the previous attempt, CoT-SelfEvolve generated the following code:
GENERATED_CODE:
```
{generated_code}
```

And it received the following feedback:
FEEDBACK:
```
{feedback}
```

In general, the steps to reason about the problem are:
- Step 1: understand the problem, what does it require?
- Step 2: analyze the GENERATED_CODE vs problem, what is the problem? does it
related with the defined problem?
- Step 3: analyze the FEEDBACK, what is the error message? what is the expected
output?

Please help the CoT-SelfEvolve agent by providing step-by-step guidance to solve
the problem. DO NOT attempt to solve the problem directly. Remember that you are
helping another agent to solve the problem, not solving the problem directly.
\end{verbatim}
\end{small}

\subsection{Example of generator CoT prompt (2)}
\label{sec:example_cot_2}
\begin{small}
\begin{verbatim}
Sure! Let's go through the steps to reason about the problem and help the CoT-
SelfEvolve agent.

Step 1: Understand the problem
The problem requires shuffling the order of the DataFrame's rows according to a
given list.

Step 2: Analyze the GENERATED_CODE vs problem
The GENERATED_CODE is incomplete and needs to be modified to solve the problem.
The `df` and `List` variables are loaded from the pickle file, but the code in-
side the `[insert]` block is missing. The goal is to shuffle the rows of the
DataFrame `df` according to the list `List`.

Step 3: Analyze the FEEDBACK
The feedback indicates that there is a NameError because the variable `result`
is not defined. The expected output is not mentioned in the feedback.

Now that we have analyzed the problem and the code, let's provide step-by-step
guidance to solve the problem.

Step 1: Shuffle the DataFrame
To shuffle the DataFrame `df` according to the list `List`, you can use the
`iloc` function of pandas DataFrame. The `iloc` function allows you to select
rows based on their integer positions. You can pass the list `List` as the argu-
ment to `iloc` to select the rows in the desired order.

Step 2: Modify the GENERATED_CODE
Replace the `[insert]` block in the GENERATED_CODE with the code to shuffle the
DataFrame. Assign the shuffled DataFrame to the variable `result`.

Step 3: Save the result
After shuffling the DataFrame, save the `result` to the pickle file.

Remember, the goal is to guide the CoT-SelfEvolve agent to solve the problem by
himself.
\end{verbatim}
\end{small}

\subsection{Correction Code Generator}
\label{sec:correction_code_generator}
\begin{small}
\begin{verbatim}
In the previous attempt, you generated the following code:
GENERATED_CODE:
```
{generated_code}
```
However, the system has given you the following instruction:
FEEDBACK:
```
{feedback}
```
Please comply with the instruction and generate the code accordingly.

To support you in solving the problem, here are the Chain-of-Thought reasoning
suggestions, you should follow these suggestions one by one, to use them as a
guide for your internal reasoning process to solve the problem.
{cot_suggestion}
\end{verbatim}
\end{small}

\end{document}